\documentclass[prl,twocolumn,showpacs,amsmath,amssymb]{revtex4-1}
\usepackage{graphicx}
\usepackage{dcolumn}
\usepackage{bm}
\def\3he{$^3$He}
\def\4he{$^4$He}

\begin{document}


\title{Mass flux characteristics in solid $^4$He for $T >$ 100 mK: Evidence for Bosonic Luttinger Liquid behavior}

\author{Ye. Vekhov}
\author{R.B. Hallock}%
\affiliation{%
Laboratory for Low Temperature Physics, Department of Physics,\\
University of Massachusetts, Amherst, MA 01003 
}%

\date{\today} 

\begin{abstract}

  At pressure $\sim$ 25.7 bar the flux, $F$, carried by  solid \4he for $T >$ 100 mK depends on the net chemical potential difference between two reservoirs in series with the solid, $\Delta \mu$, and obeys  $F \sim (\Delta \mu)^b$, where $b \approx 0.3$ is independent of temperature.
  At fixed $\Delta \mu$ the temperature dependence of the flux, $F$, can be adequately represented by $F \sim  - \ln(T/\tau)$, $\tau \approx 0.6$ K, for $0.1 \leq T \leq 0.5$ K.  A single function $F = F_0(\Delta \mu)^b\ln(T/\tau)$ fits all of the available data sets in the range 25.6 - 25.8 bar reasonably well.  We suggest that the mass flux in solid \4he  for $T > 100$ mK may have a Luttinger liquid-like behavior in this bosonic system.

\end{abstract}

\pacs{67.80.-s, 67.80.B-, 67.80.bd, 71.10.Pm}
\maketitle

Following the measurements of Kim and Chan\cite{Kim2004a,Kim2004b} and the interpretation of the possible existence of a supersolid\cite{Leggett1970}, there has been renewed interest in solid \4he.  Some have questioned the supersolid interpretation and imply that some experiments carried out to date may show no clear or only weak direct evidence for supersolid behavior\cite{Syshchenko2010,Reppy2010}.
Experiments designed to create flow in solid \4he in confined geometries by directly squeezing the solid lattice have not been successful\cite{Greywall1977,Day2005,Day2006,Rittner2009}.  We took a different approach and by creation of chemical potential differences across bulk solid samples in contact with superfluid helium have demonstrated mass transport by measuring the mass flux, $F$, through a cell filled with solid \4he\cite{Ray2008a,Ray2009b} at temperatures that extend to values above those where torsional oscillator or other experiments have focused attention.  Indeed these experiments revealed interesting temperature dependence\cite{Ray2010c,Ray2011a} in the vicinity of 80~mK, where the major changes in torsional oscillator period or shear modulus\cite{Day2007} were seen.

Here we seek to understand the behavior of $F$ for $T >$ 100 mK in more detail.  We apply a temperature difference, $\Delta T$, to create an initial chemical potential difference, $\Delta \mu_0$, between two superfluid-filled reservoirs in series with a cell filled with solid \4he. We then measure in some detail the behavior of the \4he flux through the solid-filled cell for $T > $100 mK that results from the imposed $\Delta T$ as the pressure difference between the two reservoirs changes (the fountain effect) and the chemical potential difference between the two reservoirs, $\Delta \mu$, changes from $\Delta \mu_0$ to zero. For $T > 100$ mK modest period shifts have been seen in a number of torsional oscillator experiments, in some cases even above 400 mK\cite{Penzev2008}.

\begin{figure}[b]
\resizebox{2 in}{!}{
\includegraphics{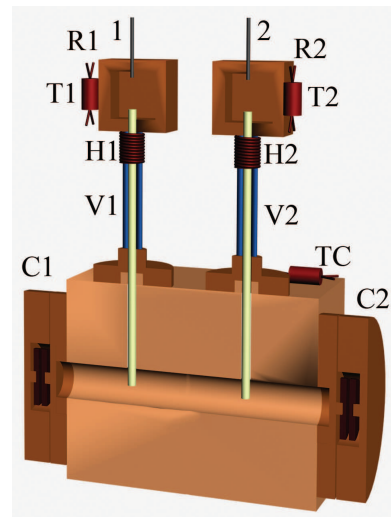}}
\caption{\label{fig:cell} (color online) 
Diagram of the apparatus.   The pressure of the solid is measured by capacitance strain gages \cite{Straty1969} C1 and C2, the pressures in the superfluid-filled reservoirs, $P1, P2$, are measured at room temperature, and the sample cell temperature is measured by thermometer TC. [Not to scale; V1 and V2 are longer than shown here.]}
\end{figure}

 Since the apparatus\cite{Ray2010c,Ray2011a} used for this work has been described in detail previously, our description here will be concise.  A temperature gradient is present across the superfluid-filled Vycor\cite{Beamish1983,Lie-zhao1986,Adams1987} rods (Figure 1), V1 and V2, which ensures that the reservoirs R1 and R2 remain filled with superfluid, while the solid-filled cell (1.84 cm$^3$) remains at a low temperature.   For the present experiments a chemical potential difference can be imposed by the creation of a temperature difference, $\Delta T =  T1 - T2 $, between the two reservoirs.  The resulting change in the fountain pressure\cite{Ray2010b} between the two reservoirs results in a mass flux through the solid-filled cell to restore equilibrium.
The
experimental protocol is designed to minimize what has been described as the ``syringe effect"\cite{Soyler2009,Ray2010a} by which sequential net injections of atoms to the cell increase the density of the solid. By a reduction in the base temperatures of R1 and R2 we also eliminate the flow restriction that would be present for too high a Vycor temperature\cite{Ray2011a}.

 To fill the cell initially, the helium gas (ultra high purity; {\it assumed} to contain $\sim$300 ppb \3he) is
 condensed through a direct-access heat-sunk capillary (not shown in Figure~\ref{fig:cell}).  To grow a solid at constant temperature from the superfluid, which is our standard technique, we begin with the pressure in the cell  just below the bulk melting pressure for \4he at the growth temperature and then add atoms simultaneously through lines 1 and 2.  Once we have created solid at the desired pressure, we close the fill lines and change the cell temperature.

\begin{figure}
\resizebox{3.5 in}{!}{
\includegraphics{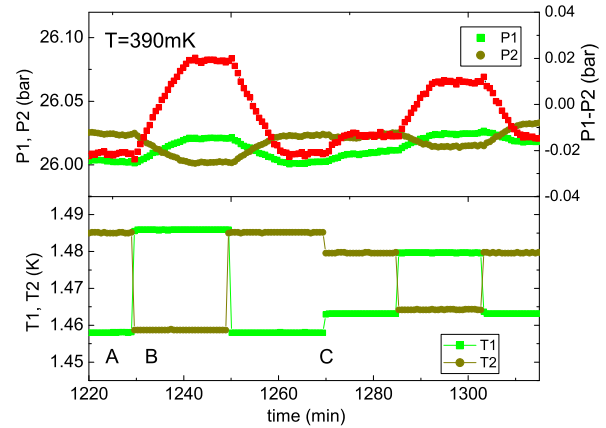}}
\caption{ \label{fig:example} (color online)  
Response of pressures $P1$ and $P2$ to the application of two different $\delta T$.  $T$ = 390 mK.  Use of heaters, H1 and H2,
results in changes in $T1$ and $T2$.  The resulting changes in $P1$ and $P2$ are
best seen as $P1-P2$, shown here (uppermost data). The small drift in $P1, P2$ of the sort seen here is typical and variable and appears to have no influence on $P1-P2$ }
\end{figure}

With stable solid \4he in the cell, we use heaters H1 (H2) to vary $T1$ ($T2$) to create chemical potential differences between the reservoirs and then measure the resulting changes\cite{Ray2010b} in the pressures $P1$ and $P2$.
 An example of the behavior seen from an application of this approach is shown in Figure \ref{fig:example}, where $P1, P2, \Delta P$ = $P1-P2$ and $T1$ and $T2$ are shown as a function of time.  A baseline reservoir temperature is first selected, $T_0$, with $T1 = T2 = T_0$. Then $T1$ is decreased by $\delta T$ while $T2$ is increased by the same interval; $\Delta T = T1 - T2 = -2\delta T$. After chemical potential equilibrium is reached (A, Figure 2), the values of $T1$ and $T2$ are interchanged (B, Figure 2); $\Delta T = T1 - T2 = +2\delta T$.
 Later $\delta T$ is changed by a small amount (C, Figure 2) and the process is continued as time elapses.  With each switch in the value of $T1 - T2$ there is a response of $P1 - P2$.  This approach is expected to create a smaller perturbation on the solid and allows us to obtain larger $\Delta \mu$ values without exceeding the upper Vycor temperature at which a significant flow limitation is encountered\cite{Ray2011a}. We take $F = d(P1-P2)/dt$ to be proportional to the flux of atoms that passes through the solid. We study $F$ as a function of $T$ and $\Delta \mu$, the chemical potential difference between R1 and R2, where $\Delta \mu = m_4[\int(dP/\rho) - \int(sdT)]$, where $m_4$ is the \4he mass, $\rho$ is the density and $s$ is the entropy per unit mass.  We report $\Delta \mu$ in units of J/g instead of J/atom.  We will report our flux values in mbar/s, where a typical value of 0.1 mbar/s corresponds to a mass flux through the cell of $\approx$ $4.8 \times 10^{-8}$ g/sec.

\begin{figure}
\resizebox{3.5 in}{!}{
\includegraphics{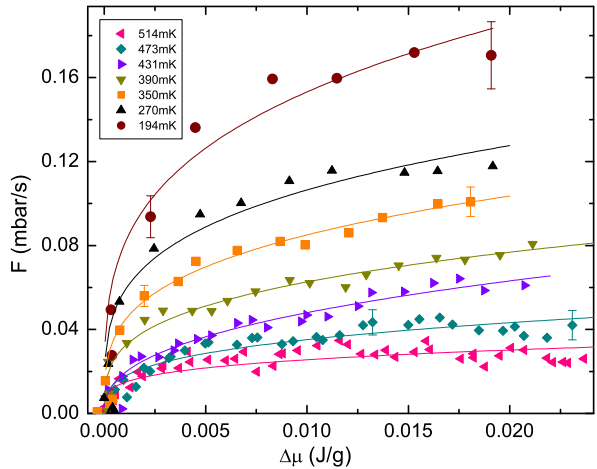}}
\caption{ \label{fig:vis} (color online) 
 Values of $F = d(P1-P2)/dt$ for
 a solid sample at a cell
pressure of $(C1 + C2)/2$ = 25.6 - 25.8~bar
 shown as a function of $\Delta \mu$ determined for the case
$\Delta T$ = 27~mK; $ \mid \Delta T \mid =  2\delta T$.  Small shifts in $\Delta \mu$ have been applied to align the data at $\Delta \mu$ = 0. The non-hysteretic flux depends on $\Delta \mu$, and can be represented by $F = A(\Delta \mu)^b$.  }
\end{figure}

\begin{figure}
\resizebox{3.3 in}{!}{
\includegraphics{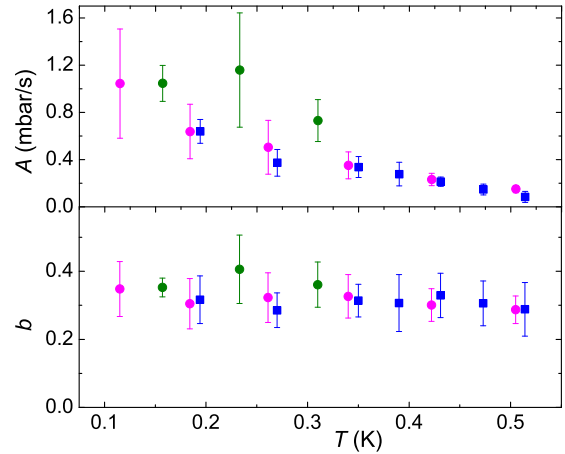}}
\caption{ \label{fig:ATBA} (color online)  
 Values of the fit parameters $A$ and $b$ as a function of temperature as determined from data including that in Figure \ref{fig:vis} (squares), obtained from a stable sample studied sequentially over several days; different symbols are for data separated by helium transfers to the apparatus. Here data that results from $ \mid \Delta T \mid $ values in the range 10 - 37.5 mK have been averaged at each $T$.   For these data sets $b$ is nearly independent of temperature with an average value $\approx$ 0.32.  }
\end{figure}

We typically consider the data in two ways: (1) $F$ as a function of  $\Delta T$ at a sequence of fixed solid \4he temperatures and (2) $F$ for fixed $\Delta T$ as a function of $T$.
For data of the first sort, we measure the dependence of the flux $F$, on the imposed temperature difference between the reservoirs R1 and R2, $\Delta T$.   Following the application of the imposed $\Delta T$ the system responds with a flux to create an increasing $\Delta P$ (the fountain effect). Thus the net chemical potential difference $\Delta \mu$ between the two reservoirs decreases to zero as $\Delta P$ increases.  Since  the flux should depend on $\Delta \mu$ we document that behavior.  An example of the relationship between $F$ and $\Delta \mu$ is shown in Figure~\ref{fig:vis} for several solid \4he temperatures. These data have error bars that are related to our ability to determine the flux from the measured  $d(P1-P2)/dt$ and this becomes more difficult at small values of $\Delta \mu$.  We find that a reasonable characterization of the data is given by $F =  A(\Delta \mu)^b$.  The results of fitting the data to this functional form for several sets of data at various cell temperatures are shown along with the data in Figure \ref{fig:vis}.  We find that $A$ has temperature dependence, but that the exponent $b$ is constant within our errors, as is illustrated in Figure \ref{fig:ATBA} for several data sets.
We conclude from the behavior seen in Figure \ref{fig:vis},
which we have seen in other samples, that our measurements here are primarily in the dissipative regime.  This dissipation may come from phase slippages, a many-body tunneling phenomena expected in a superfluid-like system.  Historically, some have explored the approach to the dissipative regime in a superfluid system by study of the flow velocity associated with a pressure gradient\cite{Kidder1962} or a decreasing gravitational pressure head\cite{Flint1974}, under quasi-isothermal conditions.  An interchange of the axes of Figure \ref{fig:vis} is reminiscent of such studies.

\begin{figure}
\resizebox{3.5 in}{!}{
\includegraphics{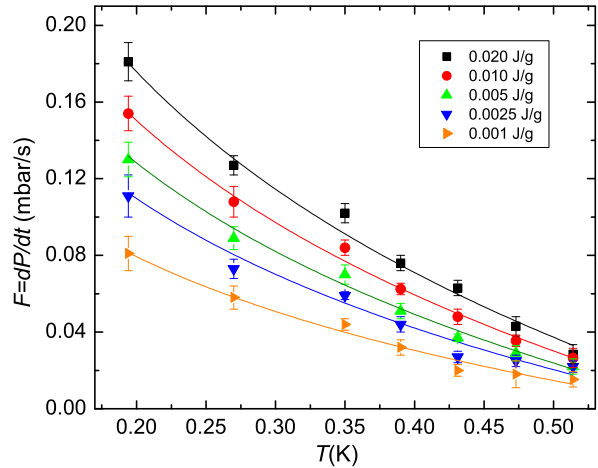}}
\caption{ \label{fig:exp} (color online) 
$F$ determined at $(C1 + C2)/2 = 25.6 - 25.8$~bar as a function of $T$ for different values of $\Delta \mu$, interpolated from the data shown in Figure 3, for the case of an applied $ \mid \Delta T \mid $ = 27 mK.
These and data for other imposed $\Delta T$ values can be represented by $F \sim -\ln(T/\tau)$.  }
\end{figure}

Next, we study $F$ {\it vs.} $T$ for several fixed values of $\Delta \mu$.  We find that the data of this sort can be reasonably well represented by a function of
the form $F \sim - \ln(T/\tau)$, where $\tau$ is a fitting parameter. As a specific example, in Figure \ref{fig:exp} we show data deduced from that shown in Figure 3  for which we find that $\tau \approx$ 630 $\pm$ 20 mK, which is consistent with earlier observations in this pressure range\cite{Ray2011a}, which showed no evidence for flux above $\approx$ 650~mK.  Typically in the temperature range studied here the behavior of the flux is reproducible and not hysteretic; for a given value of $\Delta \mu$ one can reproduce the measured $F(T)$ value (e.g. shown in Figures 3, \ref{fig:exp}) for increases or decreases in temperature.    But, at times the flux becomes unstable and can fall to values which are indistinguishable from zero. Once this happens, or the temperature is raised above $\sim 650$ mK and then lowered, it almost never is the case that a finite value of the flux will reappear when the temperature is lowered unless net atoms are added to or removed from the cell, which presumably changes the disorder in the solid.

 We find that a single function $F = F_0(\Delta \mu)^b\ln(T/\tau)$ fits the data in the range 25.6 - 25.8 bar reasonably well. From simultaneous fits to the data for the dependence on $\Delta \mu$ and $T$ for the data of Figures 3 and 5, we find average values for the parameters to be $F_0 = -0.50 \pm 0.03$ mbar/s, $b = 0.29 \pm 0.01$ and $\tau = 0.63 \pm 0.01$ K.

%

The behavior of $F$  {\it vs.}  $\Delta \mu$ appears to be qualitatively different from the behavior seen by Sasaki et al.\cite{Sasaki2006} for flow along grain boundaries on the \4he melting curve; our flux decreases with decreasing $\Delta \mu$, theirs did not.  In a saturated vapor pressure study of the decay of a superfluid \4he level, $h$, due to flow through narrow ($\sim$ 2-5 $\mu$m) slits formed between two flat plates Rorschach\cite{Rorschach1957} found that $dh/dt = a h^{1/3}$, a dependence on $h$ consistent with Gorter-Mellink friction.
Presumably $\Delta \mu \sim h$ in each case. The temperature dependence found by Rorschach closely followed the temperature dependence of the bulk superfluid density, which is distinctly different from the temperature dependence we observe, $F \sim - \ln(T/\tau)$.

Given the fact that we have a solid-filled cell off the melting curve, an unusual dependence on $T$, dissipative flux and sensitivity to disorder, we suggest a possible explanation for our observations. In confined one-dimensional geometries some authors have predicted that liquid helium might behave as a Luttinger
liquid\cite{DelMaestro2010,DelMaestro2011}. For example, in the context of liquid helium-filled carbon nanotubes, Del Maestro et al.\cite{DelMaestro2011} have recently used quantum Monte Carlo simulations to show that Luttinger liquid-like behavior should be present, with a Luttinger parameter that depends on the pore diameter. And Boninsegni et al.\cite{Boninsegni2007} have predicted that Luttinger-like behavior will be present in the cores of screw dislocations in solid helium.

In one dimension, the macroscopic behavior of bosons and fermions is the same \cite{Haldane}. A Luttinger liquid is traditionally thought of as a one-dimensional fermionic conductor in which the relaxation of current is due to the back scattering of single fermions dressed with bosonic phonon-type modes. The bosonic counterpart of this picture is the relaxation of a supercurrent due to quantum phase slippages \cite{Kane1992,Kashurnikov}. In the low temperature limit in the presence of a finite chemical potential difference, a Luttinger liquid is predicted to carry a non-Ohmic current $I $,  of the form $I \sim (\delta \mu)^p$, where $\delta \mu$ is the driving chemical potential difference, e.g. the applied voltage, and where $p$ is a constant related to the Luttinger liquid parameter, $g$.  For a single conduction channel with Luttinger liquid behavior, one expects such behavior for $k_BT/\hbar << J$, where $J$ is the flux in atoms/s.  For our work, e.g. at $T \sim 0.2$K, with $\Delta \mu \approx$ 0.01 J/g, we have a flux of $J \sim 7 \times 10^{15}$ atoms/sec.   $T \sim 0.2$K results in $k_BT/\hbar = 2.6 \times 10^{10}$.  This indicates that for Luttinger liquid behavior to be relevant to our results, the effective number of conducting channels that carry flux, $N$, should be $\lesssim 2 \times 10^5$.  Estimating the effective diameter of a channel\cite{Boninsegni2007}, this in turn indicates that the flow velocity in a given channel should be $\gtrsim 200$ cm/sec.

For a Luttinger liquid in the quantum regime where $g$ is less than unity,  and  the impedance is due to impurities,  Kane and Fisher\cite{Kane1992} predict that  $p = 2/g - 1$, independent of temperature.  The data presented in Figure~\ref{fig:ATBA}b, with $b = p \approx 0.32$, implies that $g \approx 1.52$ according to the above criterion.  On the other hand, more recent work\cite{Svistunov2012} based on an effective Hamiltonian of phase slippages suggests (for the impedance due to impurities) $p = 1/(2g - 1)$,  which implies $g \approx 2.06$.   [The result $p = 1/(2g - 1)$ is implied by the impurity correction for $g > 1$ considered by Kane and Fisher\cite{Svistunov2012}.]   If in our temperature range the flux in solid helium is carried by the superfluid cores\cite{Boninsegni2007} of edge dislocations\cite{Soyler2009}, a one-dimensional model seems relevant.    Typically one thinks of Luttinger liquids as finite-length one-dimensional systems. Here, the picture would perhaps be of a series of connected one-dimensional segments, e.g. dislocation cores.


In summary, we find that at constant solid \4he temperature the flux of atoms that pass through a cell filled with solid \4he can be reasonably represented by $F = A (\Delta \mu)^b$, with the exponent $b$ independent of temperature.  We also find that at fixed $\Delta \mu$ the temperature dependence of the flux can be rather well represented by $F = f_0\ln(T/\tau)$, consistent with the extinction of the flux above a characteristic temperature, $\tau$. We suggest that solid helium in the temperature and pressure range of this study may be an example of a Bosonic Luttinger liquid.

 We thank M.W. Ray for his previous work on the apparatus and helpful comments, B. Svistunov for a number of stimulating discussions and N. Mikhin for technical advice.  This work was supported by NSF DMR  08-55954,
   DMR 07-57701,  and by Research Trust Funds administered by the University of Massachusetts Amherst.


\bibliography{ref}

\end{document}